\documentclass[aps,prl,amsmath,amssymb,floatfix,twocolumn,amsmath,superscriptaddress,twocolumn,nofootinbib,tighten,letterpaper]{revtex4-2}
\usepackage{multirow}
\usepackage{subfigure}
\usepackage{color}
\usepackage{mathrsfs}
\usepackage[colorlinks,linkcolor=blue,citecolor=blue,urlcolor=blue]{hyperref}
\usepackage[normalem]{ulem}
\usepackage{bm}

\usepackage{amssymb}   
\usepackage{amsmath}
\renewcommand\vec[1]{\ensuremath\boldsymbol{#1}} 

\usepackage{amsfonts, relsize, color}
\usepackage{graphics}
\usepackage{graphicx}
\usepackage{subfigure}
\usepackage{hyperref}
\usepackage{color}
\usepackage{comment}

\begin{document}

\title{Emergent fractals in dirty topological crystals}

\author{Daniel J.\ Salib}
\affiliation{Department of Physics, Lehigh University, Bethlehem, Pennsylvania, 18015, USA}

\author{Bitan Roy}
\affiliation{Department of Physics, Lehigh University, Bethlehem, Pennsylvania, 18015, USA}

\date{\today}

\begin{abstract}
Non-trivial geometry of electronic Bloch states gives rise to topological insulators which are robust against sufficiently weak randomness inevitably present in any quantum material. However, increasing disorder triggers a quantum phase transition into a featureless normal insulator. As the underlying quantum critical point is approached from the topological side, small scattered droplets of normal insulators start to develop in the system and their coherent nucleation causes ultimate condensation into a trivial insulator. Unless disorder is too strong, the normal insulator accommodates disjoint tiny topological puddles. Furthermore, in the close vicinity of such a transition the emergent islands of topological and trivial insulators display spatial fractal structures, a feature that is revealed only by local topological markers. Here we showcase this (possibly) generic phenomenon that should be apposite to dirty topological crystals of any symmetry class in any dimension from the Bott index and local Chern marker for a square-lattice-based disordered Chern insulator model.        
\end{abstract}

\maketitle

\emph{Introduction}.~Thermal and quantum continuous phase transitions are fascinating phenomena in modern physics as in the vicinity of the associated critical points physical observables manifest power-law behavior in terms of universal scaling exponents~\cite{book:1, book:2, book:3, book:4, book:5}. Also, when such a critical point is approached from an ordered phase, it gradually loses spatial coherence and small incoherent islands of the disordered phase start to develop in the system. In the close proximity to the critical points, such flakes of ordered and disordered phases display fractal structures. While this phenomenon is well-appreciated in interacting systems, its jurisdiction across the disorder-controlled quantum phase transition (QPT) in noninteracting topological quantum materials remains unclear; raising the following question. \emph{How do dirty topological insulators} (TIs) \emph{become trivial or normal insulator} (NI) \emph{?}

To answer this question, we consider a specific lattice-regularized model for disordered TI and come to the following (possibly generic) conclusions that should be pertinent in dirty topological crystals from any symmetry class in any dimension. We compute a global topological invariant and the associated local topological marker in tandem. While the former one allows us to pin the critical disorder strength ($W_c$) for a topological-normal insulator QPT, the latter one provides invaluable insights into the spatial profile of local topology. We show that as a dirty TI arrives at the shore of such a QPT, small isolated droplets of a NI start to form in the system. Near the TI-NI quantum critical point (QCP), the islands of TI and NI display spatial fractal structures. We identify a few isolated tiny pockets of TI inside a NI phase, unless the disorder strength $W \gg W_c$.

\begin{figure*}[t!]
\includegraphics[width=1.00\linewidth]{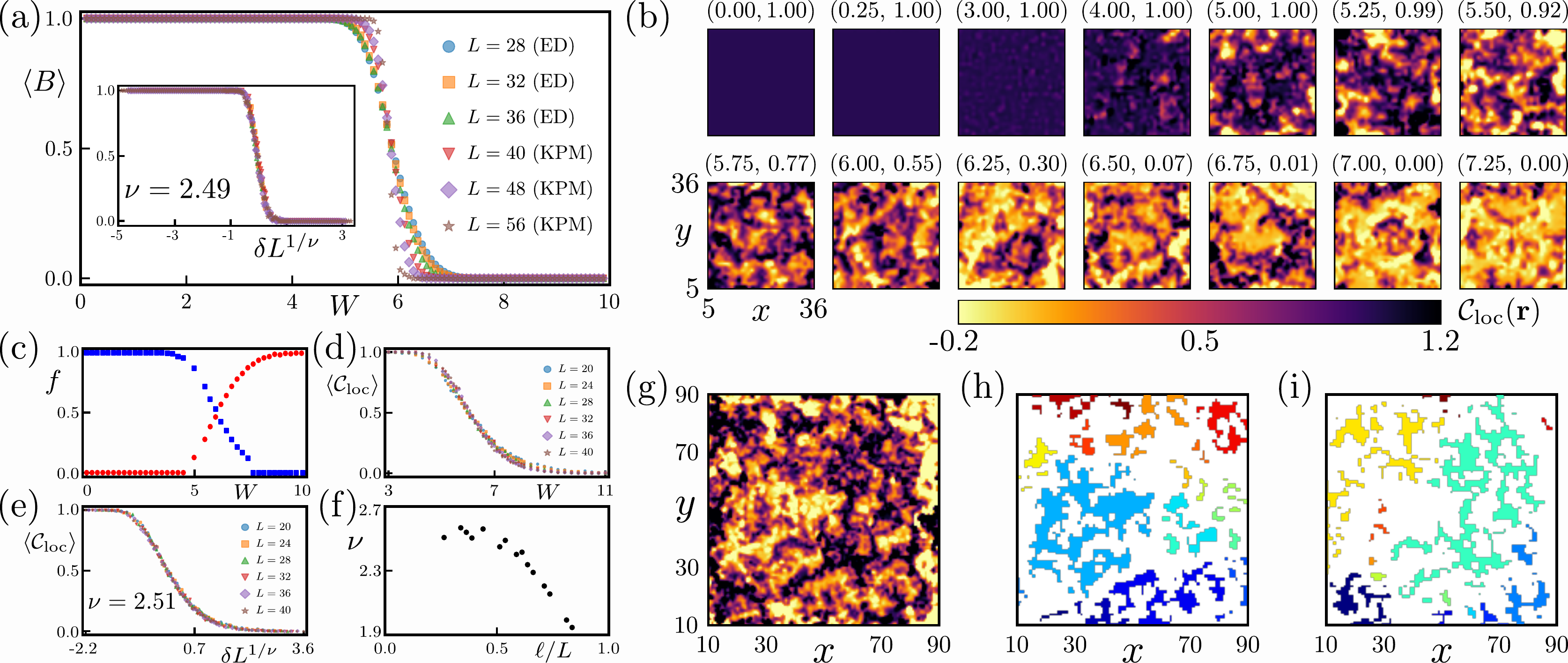}
\caption{(a) Disorder-averaged BI $\langle B \rangle$ as a function of disorder strength ($W$), obtained from exact diagonalization (ED) and kernel polynomial method (KPM), cross at $W \approx 6.0$ for different $L$ (linear dimension of square lattice), where $\langle B \rangle \approx 0.5$, marking the critical disorder $W_c$ for the TI-NI insulator QPT. Inset: Data collapse for $\delta L^{1/\nu}$ vs $\langle B \rangle$ with $\nu=2.49$, where $\delta=(W-W_c)/W_c$ and $\nu$ is the correlation length exponent. (b) Spatial distribution of the LCM $C_{\rm loc}(\vec{r})$ on a $L=40$ square lattice is shown after excluding four sites from each end of the system in both directions for a single disorder realization for each $W$. The pair of numbers correspond to $(W, \langle B \rangle)$. (c) Fraction of such an \emph{interior} area ($f$) occupied by regions with $C_{\rm loc}(\vec{r}) = 1.0 \pm 0.2$ (blue) and $0.0 \pm 0.2$ (red). (d) Disorder-averaged LCM $\langle {\mathcal C}_{\rm loc} \rangle$ in a square box of linear dimension $\ell=L/4$ at the center of the system as a function of $W$, which for all choice of $L$ cross roughly at $W=W_c$. (e) Data collapse for $\delta L^{1/\nu}$ vs $\langle {\mathcal C}_{\rm loc} \rangle$ with $\nu=2.51$ for $\ell=L/4$. {\color{blue}(f)} Variation of $\nu$ with $\ell/L$. {\color{blue}(g)} Distributions of $C_{\rm loc}(\vec{r})$ for a single disorder realization with $W=W_c$ on a $L=100$ square lattice is shown after excluding ten sites from each end of the system in both directions. Patches of {\color{blue}(h)} trivial [{\color{blue}(i)} topological] insulators with $C_{\rm loc}(\vec{r}) = 0.0 \pm 0.2$ [$1.0 \pm 0.2$] are shown in different colors, exhibiting fractal structures (see Fig.~\ref{fig:Fig2}). LCM is always computed from ED. We average over 50 (200) disorder realizations far from (near) $W_c$.    
}~\label{fig:Fig1}
\end{figure*}

\begin{figure*}[t!]
\includegraphics[width=1.00\linewidth]{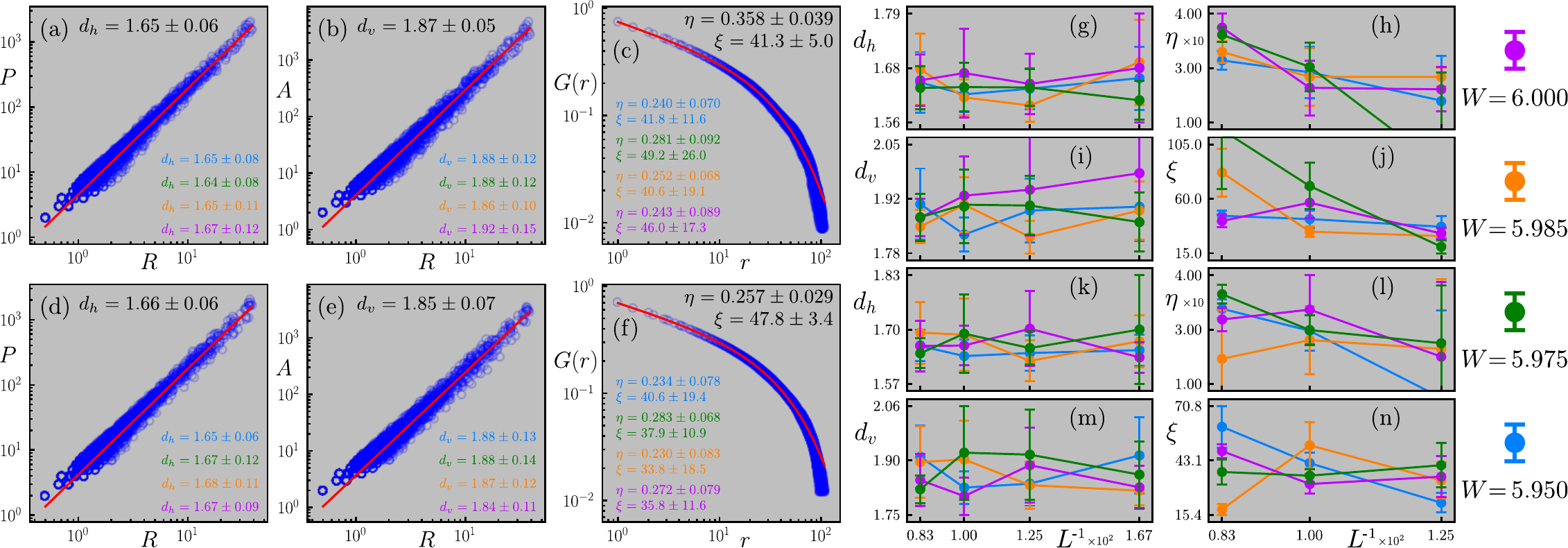}
\caption{Computation of (a) hull fractal dimension $d_h$, (b) volume fractal dimension $d_v$, and (c) anomalous dimension $\eta$ and correlation length $\xi$ from the pair-connectivity $G(r)$ at critical disorder $W=W_c=6.0$ from the islands of NIs in a $L=120$ system (in black). Panels (d)-(f) same as (a)-(c), respectively, for the islands of TIs. Dependance of (g) $d_h$, (h) $d_v$, (i) $\eta$, and (j) $\xi$ on $L$ for a few $W \lesssim W_c$ (color coded) for the islands of NIs. Panels (k)-(n) are same as (g)-(j), respectively, for the islands of TIs. We average over 50 disorder realizations. For each each $W$ we quote (color coded) the values of $d_h$ [(a) and (d)], $d_v$ [(b) and (e)], and $\eta$ and $\xi$ [(c) and (f)]. Error bars capture bootstrap standard error~\cite{SM}.   
}~\label{fig:Fig2}
\end{figure*}

As a demonstrative example, we showcase these outcomes by numerically computing the disorder-averaged Bott index (BI)~\cite{inv:1} and local Chern marker (LCM)~\cite{inv:2, inv:3, inv:4} from a square lattice-based Chern insulator model. While the BI jumps from an integer value to zero across the TI-NI QPT, the LCM displays the rich structure mentioned above. The associated fractal and anomalous dimensions are reasonably close to the ones known for the two-dimensional (2D) uncorrelated Ising-like percolation theory~\cite{Ising:1, Ising:2, Ising:3, Ising:4, Ising:5}; slight deviations from which stem from the unavoidable spread of the LCM around the integer and trivial values in disordered systems. See Figs.~\ref{fig:Fig1} and~\ref{fig:Fig2}. Therefore, our theoretical predictions unify emergent fractal phenomenon near critical points that so far has been studied in interacting, non-topological disordered, and statistical systems by introducing a new member in this family: dirty topological crystals.

\emph{Model}.~The minimal Hamiltonian for 2D Chern insulators of spinless or spin-polarized fermions is $h=\mathbf{d}(\mathbf{k}) \cdot {\boldsymbol \tau}$. The vector Pauli matrix $\bm{\tau}=(\tau_x,\tau_y, \tau_z)$ operates on orbitals with parity eigenvalues $\tau=\pm$. We choose ${\mathrm d}_1(\mathbf{k})=t \sin(k_x a)$, ${\mathrm d}_2(\mathbf{k})=t \sin(k_y a)$, and ${\mathrm d}_3(\mathbf{k})=m_0 - t_0 [\cos(k_x a)+\cos(k_y a)]$~\cite{QWZ}. Here, $a$ is the lattice constant, $t$ ($t_0$) is the hopping amplitude between the opposite (same) parity orbitals living on the nearest-neighbor sites of a square lattice, and $m_0$ denotes on-site staggered potential, yielding a uniform Dirac mass. This system supports TI (NI) for $|m_0/t_0|<2$ ($|m_0/t_0|>2$). We set $t=t_0=m_0=1$, yielding a TI with the band inversion at the $\Gamma$ point, characterized by the first Chern number $C=1$ for the filled valence band, obtained by integrating its Berry curvature over the first Brillouin zone~\cite{TKNN}. We disregard particle-hole asymmetry as it does not affect the topology of an insulating system.

\emph{Disorder}.~In the presence of disorder, the translational symmetry gets broken and the notion of a Bloch Hamiltonian becomes moot. We are then forced to compute the associated topological invariant by diagonalizing the corresponding real space tight-binding Hamiltonian, which on a square lattice for the chosen $\mathbf{d}(\mathbf{k})$ vector reads as 
\allowdisplaybreaks[4]
\begin{eqnarray}~\label{eq:realspaceTB}
H^{\rm SL}_{\rm TB} &=& \sum_{\vec{r}} \bigg[ m_0 \Psi^{\dagger}_{\vec{r}} \tau_z \Psi_{\vec{r}}
+  \sum_{j=x,y} \bigg\{ \frac{t}{2i} \Psi^{\dagger}_{\vec{r}} \tau_j \Psi_{\vec{r}+\hat{e}_j} \nonumber \\
&+& \frac{t_0}{2} \Psi^{\dagger}_{\vec{r}} \tau_z \Psi_{\vec{r}+\hat{e}_j} \bigg\} \bigg]
+ \sum_{\vec{r}} \Psi^{\dagger}_{\vec{r}} \tau_0 V(\vec{r}) \Psi_{\vec{r}}.
\end{eqnarray}
Here, $\hat{e}_j=a \hat{j}$ with $\hat{j}$ as the unit vector along $j=x,y$, $\Psi^\top_{\vec{r}}=(c_{\vec{r},+},c_{\vec{r},-})$, and $c_{\vec{r},\tau}$ is the fermionic annihilation operator at position $\vec{r}$ with parity $\tau$. The last term encodes on-site potential disorder, the dominant source of elastic scattering in real materials. On each lattice site, we sample $V(\vec{r})$ uniformly and randomly from a box distribution $[-W/2,W/2]$, where $W$ is the disorder strength. We denote the corresponding matrix operator as ${\boldsymbol W}$ with its elements given by ${\boldsymbol W}_{i,j} = w_{i}\delta_{ij}$, where $i$ and $j$ are site indices, $w_{i} \in [-W/2,W/2]$, and $\delta_{ij}$ is the Kronecker delta symbol. While ${\boldsymbol W}$ is traceless in sufficiently large systems for each disorder realization, in order to ensure this property that minimizes the shift in the Fermi level in moderate systems we replace ${\boldsymbol W}$ by ${\boldsymbol W}-\delta_W {\boldsymbol I}_N$ in numerical calculations. Here, $\delta_W={\rm Tr}({\boldsymbol W})/N$ is a \emph{small} constant, $N$ is the total number of sites in the system, and ${\boldsymbol I}_N$ is an $N$-dimensional identity matrix.

\emph{Topological invariants}.~In clean systems, the BI ($B$), computed upon diagonalizing $H^{\rm SL}_{\rm TB}$ [Eq.~\eqref{eq:realspaceTB}], enjoys a one-to-one correspondence with the first Chern number ($C$); $C \equiv B$. In dirty systems, we compute the disorder-averaged Bott index $\langle B \rangle$ to underpin the TI-NI QPT. We assume that the system is at half-filling and zero temperature, and define the projection operator $\mathcal{P}$ onto the valence band and its complementary operator $\mathcal{Q}$ as  
\allowdisplaybreaks[4]
\begin{equation}\label{eq:projector}
    \mathcal{P}=\sum_{n=1}^{N}\vert{n}\rangle\langle{n}\vert \:\:\: \text{and} \:\:\: \mathcal{Q} = {\boldsymbol I}_N - \mathcal{P},
\end{equation}
respectively. Here $\vert n \rangle$ are the algebraically ordered eigenstates of $H^{\rm SL}_{\rm TB}$. A unitary phase operator in the $x$ direction is defined as ${\boldsymbol U}_x = \exp{(2\pi{i}{\boldsymbol X} \otimes \tau_0/L_x)}$, where $\otimes$ stands for the Kronecker product and the matrix elements of the position operator are ${\boldsymbol X}_{i,j}=x_i \delta_{ij}$ with $x_i\in[1, L_x]$. The linear dimension of the system in the $x$ direction is $L_x$. Notice that ${\boldsymbol U}_x \vert x^\tau_i \rangle = \exp{(i2\pi x_i/L_x)}\vert x^\tau_i \rangle$. Here, $\vert x^\tau_i \rangle$ is the site-localized Wannier state at $x_i$ on an orbital with parity eigenvalue $\tau$. Similarly, we define a unitary phase operator ${\boldsymbol U}_y$ in the $y$ direction. Throughout we take $L_x=L_y=L$ (say). The projections of these phase operators onto the filled states are ${\boldsymbol V}_q = \mathcal{Q} + \mathcal{P} {\boldsymbol U}_q\mathcal{P}$ for $q=x$ and $y$. Then the net phase around closed loop constructed with ${\boldsymbol V}_x$ and ${\boldsymbol V}_y$ yields~\cite{inv:1} 
\begin{equation}~\label{eq:bott_index}
    B = \frac{1}{2\pi}\text{Im}[\text{Tr}[\ln[ {\boldsymbol V}_x {\boldsymbol V}_y {\boldsymbol V}_x^\dagger {\boldsymbol V}_y^\dagger]]].
\end{equation}
To respect the periodic nature of the phase operators, we always compute the BI on a square lattice with periodic boundary conditions.

Naturally, unveiling the spatial variation of topology is beyond the scope of BI, for which we compute the LCM at every site, encoded in the operator $C_{\rm LCM} = 4\pi\text{Im}\left[\mathcal{P} {\boldsymbol X} \mathcal{Q} {\boldsymbol Y} \mathcal{P} \right]$. On a site at $\vec{r}$ the LCM is~\cite{inv:2, inv:3, inv:4} 
\begin{equation}\label{eq:local_chern_number}
    \mathcal{C}_{\rm loc}(\mathbf{r}) = \sum_{\tau=\pm} \langle\mathbf{r}, \tau \vert{C_L}\vert \mathbf{r}, \tau\rangle.
\end{equation}
The LCM is insensitive to the boundary conditions, as ${\boldsymbol X}$ and ${\boldsymbol Y}$ are not periodic. The LCM near the boundary of the system deviates considerably from the bulk quantized value even in clean systems~\cite{inv:2}. Thus, we always exclude several sites near the boundaries of the square lattice while displaying and analyzing LCM.

Usually, we rely on exact diagonalization (ED) to compute ${\mathcal P}$, which becomes time consuming in large systems when in addition we need to perform disorder averaging. This limitation can be circumvented by computing ${\mathcal P}$ using the kernel polynomial method (KPM)~\cite{KPM:RMP, KPM:PRR}, detailed in the Supplemental Material (SM)~\cite{SM}. However, KPM does not count and order the eigenstates while constructing ${\mathcal P}$. In KPM, the eigenvalue spectrum can only be truncated at a specified energy value. Therefore, we work under the assumption that the numbers of states below and above the Fermi energy ($E_F=0$) remain equal even in the presence of disorder, which is typically the case when disorder configuration is set by the matrix ${\boldsymbol W}-(\delta_W) {\boldsymbol I}_N$. As the BI can only take integer values, a slight deviation from this assumption in KPM averages out and we obtain the same results for $\langle B \rangle$ as from ED. However, no such constraint applies for the LCM, for which we always rely on ED.

\emph{Phase diagram}.~First, we compute $\langle B \rangle$ using ED (for smaller $L$) and KPM (for larger $L$) as a function of $W$, see Fig.~\ref{fig:Fig1}(a). In the weak (strong) disorder regime the system describes a TI (NI) with $\langle B \rangle=1.0$ ($0.0$). The curves for $W$ vs $\langle B \rangle$ for different $L$ cross around $W \approx 6.0$, where $\langle B \rangle \approx 0.5$, defining the critical disorder $W_c$ for the TI-NI QPT. With a single-parameter scaling ansatz $\langle B \rangle= F(\delta L^{1/\nu})$, where $F$ is an unknown universal function of its argument, $\delta=(W-W_c)/W_c$, and $\nu$ is the correlation length exponent, determining the universality class of this QPT, we find best quality data collapse for $\nu=2.49$, close to its current estimation for the quantum Hall plateau transition~\cite{QHPT:1, QHPT:2, QHPT:3, QHPT:4, QHPT:5, QHPT:6}, see inset of Fig.~\ref{fig:Fig1}(a).

Such a seemingly featureless QPT encodes fascinating rich structure, revealed by the LCM, see Fig.~\ref{fig:Fig1}(b). In the clean and weakly disordered systems $C_{\rm loc}(\vec{r}) \approx 1$ in the entire interior of the system, when $\langle B \rangle=1$. With increasing disorder isolated droplets of NIs where $C_{\rm loc}(\vec{r}) \approx 0$ start to form in the system even inside the TI phase with $\langle B \rangle=1$. As the system approaches the TI-NI QCP, the droplets of topological and normal insulators occupy almost equal area of the system when $\langle B \rangle = 0.5$, see Fig.~\ref{fig:Fig1}(c). By the same token, a NI with $\langle B \rangle = 0$ fosters a few isolated small flakes of TIs when $W \gtrsim W_c$, which completely disappear only for $W \gg W_c$.

The disorder-averaged LCM $\langle {\mathcal C}_{\rm loc} \rangle$, averaged over the sites of a square box of linear dimension $\ell=L/4$, placed near the center of the system, display a similar behavior as $\langle B \rangle$, see Fig.~\ref{fig:Fig1}(d). With the scaling ansatz $\langle {\mathcal C}_{\rm loc} \rangle = G(\delta L^{1/\nu},\ell/L)$, where $G$ is another universal function of $\delta L^{1/\nu}$ when $\ell/L \ll 1$, we obtain best quality data collapse for $\nu=2.51$, see Fig.~\ref{fig:Fig1}(e). Fig.~\ref{fig:Fig1}(f) shows that only for $0.2 \lesssim \ell/L \lesssim 0.4$, the values of $\nu$ obtained from the data collapses of $\langle B \rangle$ and $\langle {\mathcal C}_{\rm loc} \rangle$ are sufficiently close.

Since the value of $C_{\rm loc}(\vec{r})$ is not strictly restricted to any integer value, especially in the presence of disorder, the islands of TI and NI in dirty topological crystals are identified with $C_{\rm loc}(\vec{r}) = 1.0 \pm 0.2$ and $C_{\rm loc}(\vec{r}) = 0.0 \pm 0.2$, respectively, in Figs.~\ref{fig:Fig1}(g). In the close proximity to the TI-NI QCP, individual droplets of NI and TI assume irregular fractal structure, as shown in Figs.~\ref{fig:Fig1}(h) and~\ref{fig:Fig1}(i), respectively. Next, we quantitatively establish such \emph{emergent} fractal structures near the TI-NI QPT. 

\begin{figure}[t!]
\includegraphics[width=1.00\linewidth]{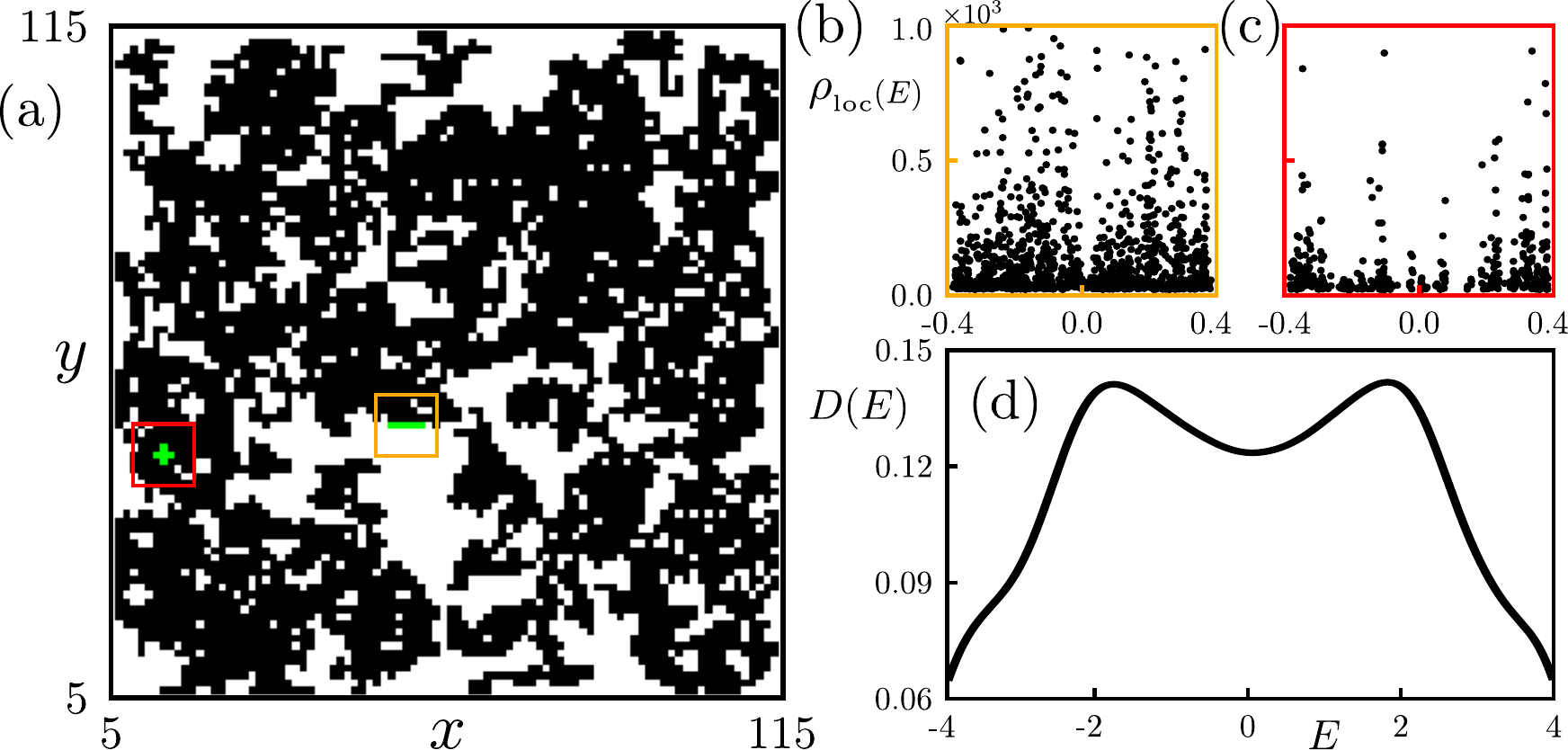}
\caption{Distribution of the LCM inside a $L=120$ square lattice for a single disorder realization of strenght $W=5.25$ showing the topological (trivial) regions with $C_{\rm loc}(\vec{r}) \approx 1.0$ ($0.0$) in black (white). Local density of states $\rho_{\rm loc}(E)$ (b) at the edge and (c) in the interior of the topological region, computed over five green colored sites within the golden and red boxes, respectively, showing gapless and multi-gapped structures near the Fermi energy $E_F=0$. (d) Corresponding average density of state $D(E)$ as a function of energy $E$.       
}~\label{fig:Fig3}
\end{figure}

\emph{Fractal analysis}.~Analysis of the fractal clusters begins with the identification and labeling of individual ones. For each fractal, we compute three geometric quantities; the radius of gyration ($R$), perimeter ($P$), and area ($A$). Here we mention the key steps, while relegating details to the SM~\cite{SM}. Since the fractals are embedded in two dimensions, the number of sites within it yields $A$, the number of sites bordering the interior of the cluster determines $P$, and $R$ is the root mean square distance of each site at $\mathbf{r}_i$ from the geometric center of the fractal at $\mathbf{r}^c$, given by $R=[R^2_1 + R^2_2 \cdots + R^2_A]^{1/2}$, where $R_i=\left| \mathbf{r}_i - \mathbf{r}^c \right|/\sqrt{A}$. The universal and scale-independent (when $L \gg R$) hull ($d_h$) and volume ($d_v$) fractal dimensions are given by $P \sim R^{d_h}$ and $A \sim R^{d_v}$, respectively. The pair connectivity $G(r)$, measuring the probability of a given pair of sites separated by distance $r$ belonging to the same cluster, in general feature both power-law and exponential decays with the scaling form $G(r) \sim \left( r/\xi \right)^{-(d - 2 + \eta)} \exp\left( -r/\xi \right)$. Here, $d=2$ is the embedding dimension, $\eta$ is the anomalous dimension, and $\xi$ is the correlation length, representing the average scale over which two points remain connected with a given cluster. At the QCP, the distribution of the LCM ideally forms two large cohesive fractals spanning the entire system and thus $\xi \approx L$ therein. At this point the exponential decay of $G(r)$ gets suppressed, thereby featuring a power-law decay, determined by $\eta$. These scaling forms capture how clusters fill the space in a self-similar manner, deviating from integer dimensions, near $W=W_c$.

Such an analysis is explicit shown on a $L=120$ system for $W=6.0$ and for the clusters of both TI and NI in Fig.~\ref{fig:Fig2}, where we also display the $L$-dependence of $d_h$, $d_v$, $\eta$, and $\xi$ for a few values of $W \lesssim W_c$. When averaged over $L$, the values of these quantities are reasonably insensitive to $W$ near $W_c$ and are close, yet slightly far from the ones for the 2D uncorrelated Ising-like percolation theory for which $d_h=1.75$, $d_v=1.9$, and $\eta=0.207$~\cite{Ising:1, Ising:2, Ising:3, Ising:4, Ising:5}. In particular for $W=6.0$, we obtain (after averaging the values over various $L$) $d_h=1.67 \pm 0.12 \; (1.67 \pm 0.09)$, $d_v=1.92 \pm 0.15 \; (1.84 \pm 0.11)$, $\eta=0.243 \pm 0.089 \; (0.272 \pm 0.079)$, and $\xi=46.0 \pm 17.3 \; (35.8 \pm 11.6)$ for the fractal droplets of NI (TI). Such mismatches stem from the fact that the LCM in dirty systems spreads considerably around expected integer and trivial values, introducing an unavoidable ambiguity in properly identifying the sites belonging to specific fractal clusters of NI and TI, while Ising spins can only have two orientations. Otherwise, a correspondence between these two systems can be established as follows. Two possible distributions of the LCM, centered around $0$ and $1$, map onto two projections of spin. And with increasing disorder randomly-placed droplets of NI fluid with ${\mathcal C}_{\rm loc}(\mathbf{r}) \approx 0.0$ \emph{percolates} through the large cluster of TI with ${\mathcal C}_{\rm loc}(\mathbf{r}) \approx 1.0$.

To establish the universal nature of emergent fractal droplets of local TIs and NIs, we repeat the entire analysis, detailed in the SM~\cite{SM}, for $t=t_0=1.0$ and $m_0=1.5$ for which the band gap is half of that with $m_0=1.0$. In this case $W_c=6.70$ around which we find $d_h=1.665 \pm 0.084$ ($1.646 \pm 0.052$), $d_v=1.882 \pm 0.068$ ($1.850 \pm 0.069$), $\eta=0.277 \pm 0.028$ ($0.241 \pm 0.086$), and $\xi=62.8 \pm 13.4$ ($55.3 \pm 14.0$) from the fractal islands of NIs (TIs), after averaging the values over various $L$. Hence, emergent fractal islands of TIs and NIs share the same universal behavior irrespective of the band gap, and in two dimensions they are always similar to the uncorrelated 2D Ising-like percolation theory. Although, existing numerical results strongly suggest that these conclusions should hold sufficiently close to the band gap closing point at $m_0=2.0$, the corresponding numerical analysis must be performed in much larger systems to account for finite-size effects in the presence of a small band gap. The requisite computation time then grows very rapidly.

\emph{Discussions}.~We show that as a precursor of the disorder-induced TI-NI QPT, isolated small droplets of incipient NI start to nucleate inside the TI phase. In the close proximity to the associated QCP, droplets of TI and NI display fractal structures. The corresponding fractal and anomalous dimensions are reasonably close to the ones for the 2D Ising-like percolation theory. The predicted fractal structure in the local topology should be germane to dirty topological crystals of any Altland-Zirnbauer symmetry class and dimensionality~\cite{AZ:1, AZ:2, AZ:3, AZ:4}, as for all of them local topological markers can be computed~\cite{inv:4}, including topological superconductors~\cite{inv:5}. The emergent fractal structures of the ground state, manifesting via local irregular droplets of TIs and NIs, near the disorder-driven QCP, separating two topologically distinct insulators, are distinct from the fractal-type structure of a few isolated critical wavefunctions residing near the Fermi energy, previously reported for Anderson~\cite{fractalwavefunction:1, fractalwavefunction:2} and disordered topological~\cite{fractalwavefunction:3, fractalwavefunction:4} models.

As the fractal and anomalous dimensions are sufficiently close for the clusters of TI and NI, the predicted fractal structure in real materials can be mapped from the local density of states (LDOS), measurable via scanning tunneling microscopy (STM)~\cite{STM:1, Ising:5}, by solely focusing on the islands of TI. Such measurements should feature multi-gapped (gapless) LDOS spectrum near the Fermi energy inside (along boundaries of) the topological islands [see Fig.~\ref{fig:Fig3}], falling within the spatial resolution window of STM measurements (a few $\mathrm{\AA}$). The fact that a gapped (gapless) spectrum is observed in the interior (at the edges) of a cluster, as shown in Fig.~\ref{fig:Fig3}, justifies the working assumption that regions with LCM $C= 1.0 \pm 0.2$ and $0.0 \pm 0.2$ represent local topological and trivial insulators, respectively. Gapless edge modes residing along the boundaries between topologically distinct islands can also be responsible for the \emph{enhancement} of electrical and thermal Hall and longitudinal conductivities near the topological QPTs, as reported in recent numerical works~\cite{numericsSKD:1, numericsSKD:2}. The STM measurements can be performed on available quantum anomalous Hall or Chern insulators, realized thin films of Bi$_2$Se$_3$, Bi$_2$Te$_3$, and Sb$_2$Te$_3$ doped by Cr or V or Fe, for example~\cite{QAHI:1, QAHI:2, QAHI:3}, and quantum spin Hall insulators in CdTe-HgTe~\cite{QSHI:1, QSHI:2} and InAs-SbTe~\cite{QSHI:3} quantum wells by tuning the disorder strength therein. Recently the lattice model for the Chern insulator~\cite{QWZ} has been engineered on optical lattices~\cite{OL:1}, where disorder strength can be tuned in a controlled fashion and LDOS can be measured using local radio frequency spectroscopy~\cite{OL:2, OL:3, OL:4}, to identify the proposed emergent fractal structures near the TI-NI QPT.

\emph{Acknowledgments}.~This work was supported by the NSF CAREER Grant No.\ DMR-2238679 of B.R. 

\emph{Data availability}.~Numerical codes developed and data generated in this work are available via Zenodo~\cite{ZenodoFractals}.


\end{document}